\documentclass[authoryear,12pt]{elsarticle}

\usepackage{mathtext,amssymb}
\usepackage{epsfig}
\usepackage{graphics}

\newcommand{\magdot}[1]{#1\!\!\ensuremath{^{\rm m}}}

\newcommand{\Msun}{\ensuremath{\rm M_\odot}}
\newcommand{\Lsun}{\ensuremath{\rm L_\odot}}
\newcommand{\Msunyr}{\ensuremath{\rm M_\odot \, yr^{-1}}}

\newcommand{\kms}{\ensuremath{\rm km \, s^{-1}}}
\newcommand{\ergl}{\ensuremath{\rm erg \, s^{-1}}}
\newcommand{\yr}{\ensuremath{\rm yr}}
\newcommand{\si}{\mathop{\mathrm{si}}\nolimits}
\newcommand{\ci}{\mathop{\mathrm{ci}}\nolimits}

\journal{New Astronomy}

\begin{document}

\begin{frontmatter}

\title{Stochastic Variability of Luminous Blue Variables}
\author{P. Abolmasov}
% \author{?? A. V. Zharova}

\address{Sternberg Astronomical Institute, Moscow State University,
  Moscow, Russia 119992; Email: \emph{pavel.abolmasov@gmail.com}}
%  Tel.: +7 (495) 9392046;  Fax: +7 (495) 9328841}

% \maketitle

%\label{firstpage}

\begin{abstract}
Using the archives of the American Association of Variable Stars
Observers and our own data, we analyse the long-term variability of
several well-studied Luminous Blue Variables (LBVs) aiming on a
general picture of stochastic variability of these objects. The
power density spectra of all the selected objects may be generally described by
a single power law contaminated by observational noise at higher
frequencies.  The slopes of the power-law component are close to
$p=2$ (where $PDS \propto f^{-p}$, and $f$ is frequency) for strongly
variable flaring objects like AG~Car and significantly smaller ($p\sim
1.3$) for P~Cyg where brightness variation amplitude is $\lesssim
1\magdot{\,}$ and dominated by slow low-amplitude
variability. 
The slope holds for about two orders of magnitude in the
frequency domain, though peaks and curvatures are present at 
$f\simeq 10^{-2}\div 10^{-3}\rm d^{-1}$. 
We show that pseudo-photosphere
approach to variability may explain the power-law shape of the
variability spectrum at higher frequencies. 
However, the observed spectra are
actually rather ``red'' than ``brown'': flux variations are correlated up
to tens of years that is much longer than the characteristic
refreshment time scales of the pseudo-photosphere. We propose
that several stochastic noise components produce the power spectra of
LBVs. 
%We propose a mechanism able to reproduce the steep slopes
%of the flaring objects and the flickering spectrum of P~Cyg.
\end{abstract}


\end{frontmatter}

\section{Introduction}\label{sec:intro}

Flares from luminous blue variables (LBV) 
are studied for about four hundred years, since the 1600
flare of P~Cyg \citep{degroot}. However, these objects are rare and thus
relatively unstudied.
%The variability class itself is refer S~Dor
%variables of the Magellanic Clouds, Hubble-Sandage
%variables in nearby galaxies \citep{HS53} and  \citep{conti84}. 
As variable objects, LBV are characterised by a complex hierarchy of
variability timescales and amplitudes \citep{HD94}. Usually, three
empirical variability scales are distinguished:

\begin{enumerate}

\item microvariations, with amplitudes $\sim 0\magdot{.}1$ and
  relatively small
  variability timescales, from days to months

\item  S~Dor cycles, or eruptions (following \citet{HD94}) with
  amplitudes $\sim 1\div 2\magdot{\,}$, years to
  tens of years in length

\item giant eruptions with similar and larger (up to $\sim 100\yr$)
  timescales and larger amplitudes ($\gtrsim 2\magdot{\,}$)
\end{enumerate}

Physically, giant eruptions may differ from the variations of the
second type by significant changes in the bolometric luminosity (see
discussion in section \ref{sec:disc:lvar}). 
Some authors like \citet{genderen_perceptions} propose division of
the second type variability into normal and very long timescale S~Dor
cycles. Longer-timescale variations indeed seem to represent a
separate mode of LBV variability, that is most evident in the
XXth-century light curve of P~Cyg (see next section) 
that is free from ordinary flares but dominated by
microvariations superimposed over slow variations $\gtrsim 10 \rm yr$
in length. 

%Period searches for LBV stars are seldom successful, either due to
%the real lack of periodicities or because of strong interference from
%the stochastic component. However, s
Several well-studied LBVs 
exhibit periodic variability components. 
In particular, \citet{genderen_new} mention a $\sim 1 \rm yr $
period for AG~Car and a $\sim 7\rm yr$ period for S~Dor. Probably, these
periods do not correspond to any coherent process but
rather reflect some characteristic timescales close to the duration
timescale of a single flare. 

$\eta$ Carinae is an exceptional case in this sense. The well-known
5-year period is stable on the timescales of tens of years that
supports the idea that this object is indeed a (rather broad)
binary \citep{damineli}. 
Note however that the orbital separation of a massive binary with a 5-year
period should be $a\simeq 2\times 10^{14} (M/100\Msun )^{1/3}
(T/5\yr)^{2/3} \rm cm$. Hydrostatic radii for LBV and related stars
are $\sim 10^{12}\rm cm$ that makes the expected effects of binary
interaction on the central machine (-s) of $\eta$~Car very small. In
this study, we find the broad-band PDS of $\eta$~Car quite similar to
that of other flaring LBVs such as AG~Car.

%\bigskip

All the observed variability time scales of LBV stars lie between their
dynamic and thermal times.
Dynamic time scale is determined by the mean density of the
stellar hydrostatic core. Outflowing matter can not take part in
pulsations because the wind rapidly becomes
supersonic. Straightforward estimate for the dynamic timescale as the
free-fall time yields:

\begin{equation}\label{E:tdyn}
t_{dyn} \simeq 0.6 \left( \frac{R_\star}{10^{12}\rm cm}\right)^{3/2}
\left( \frac{M_\star}{100 \Msun}\right)^{-1/2} \rm d
\end{equation}

Due to period doubling (see for example \citet{buchler93}) and instability of
pulsational modes, lower-frequency modes are excited during
pulsations. Probably, this effect is observed in ``ordinary'' hot
supergiant stars, where microvariations occur at unexpectedly long
time scales of several days \citep{genderen_OBA}. Flaring LBV
variability is still much longer.

On the other hand, Kelvin-Helmholtz
time scale for a typical hypergiant is:

\begin{equation}\label{E:tKH}
t_{KH} \sim \frac{GM}{RL} \simeq 2\times 10^4 \frac{M_\star}{100
  \Msun} \left( \frac{R_\star}{10^{12}\rm cm}\right)^{-1}
\left( \frac{L_\star}{10^{6}\Lsun}\right)^{-1} \rm yr
\end{equation}

Unfortunately, these time scales are unreachable for modern
observations, but may be, in principle, 
studied indirectly by observations of the
ejecta, possibly light echoes from strong flares and using the
statistical properties of LBV ensembles. 

All the considered
variability scales are somewhere in between and may correspond either
to secular effects in dynamic-timescale evolution (``slow pulsations'')
or to effects in the expanding atmosphere that has, evidently, longer
characteristic timescales than the hydrostatic core. Stellar wind, ascribed some
velocity $v_w$, mass loss rate $\dot{M}$ and opacity $\kappa$, is
characterised by the radius of the (pseudo-) photosphere $R_{ph} \simeq \kappa
\dot{M}/ 4\pi v_w$ and corresponding time scale that has the physical
meaning of the replenishment time for the matter in the wind:

\begin{equation}\label{E:tref}
t_{ref} \sim R_{ph} / v_w \simeq 2 \frac{\kappa}{\kappa_T}
\frac{\dot{M}}{10^{-5} \Msunyr} \left( \frac{v_w}{100\kms}\right)^{-2} \rm d
\end{equation}

Here, $\kappa_T \simeq 0.4 \rm cm^2 g^{-1}$ is % has the physical meaning of
the Thomson cross-section for the scattering by free electrons in a
hydrogen-rich gas. Real
opacity is higher and contains some contribution of true absorption
processes that may increase the characteristic time scales by about an order
of magnitude (see, for example, \citet{OPAL}, bearing in mind
that, for a typical LBV, $\lg T \sim 4\div 5 $ and $\lg
R \sim -5\div -7$ for the inner wind).
Despite the apparently small value of $t_{ref}$, it is an important
estimate for the characteristic variability time of an optically thick
 hot stellar wind. 
%, variations in the
%photosphere radius should be tightly
%connected to LBV variability \citep{lamers87}. 
%especially to the S~Dor cycles that
%conserve the bolometric luminosity but influence the observed visual
%magnitudes through bolometric corrections.
 Longer time scales arise
due to higher opacities
Mass loss modulation during strong outbursts leads to even longer time
scales, up to years.
% during flares, but also because the
%mass loss rate and wind
%velocity are strongly modulated leading to the time scales $\gtrsim 1\yr$
%during flares increasing with the amplitude of the flare. 

% However, it should be noted that b
Bolometric luminosity is not
conserved during strong flares (see discussion in section
\ref{sec:disc:lvar}). Besides this, modeling
 LBV photospheres provides evidence  \citep{koter}
for more complicated nature of
the variability of these objects, incorporating pulsations, mass loss
variations and some additional mechanisms (such as unstable pulsational
modes) leading to tremendous energy release during giant eruptions. 

% \bigskip

%Due to some reason, t
The overall power density spectrum (PDS) of LBV
variability has not been considered as a whole. It is tempting to
compare LBV stars to % their variations to those of
 active galactic nuclei where different time scales
of a single flicker noise were for a long time interpreted as
physically distinct variability processes (see for example
\citet{terebizh4151}). Below we show that
indeed for several LBV stars, broad-band power spectra have steep
power-law shapes.
% The reason here is probably not in the presence of a
%single variability mechanism but rather in existance of several kinds
%of noise responsible for ``red'' and ``brown'' spectra in different
%frequency ranges. 

In this work we aim for the broad-band PDS of LBV variability. In
the next section, we make a compilation of observational data on
several reasonably well-studied LBV stars and present their PDS in the
following section \ref{sec:AR}. A possible explanation for the observed
broad-band PDS shapes is proposed in section \ref{sec:varshape}. Results
are discussed in section \ref{sec:disc}.

\section{Observational Data}\label{sec:obs}

We selected three well-studied and representative Galactic LBV stars. 
For these objects (see table \ref{tab:objs}),
relatively long observational series exist, spanning periods of time
in excess of 50 years. The
data were taken from the archives of the American Association for Variable
Star Observers. Earlier data are primarily visual magnitude estimates,
therefore we use only visual magnitudes. We also checked that the
data series do not have gaps longer than several months and binned the
observational data points by five to diminish the effect of single
erroneous estimates and the round-off effect connected to the low
precision ($\sim 0\magdot{.}1$) of visual magnitude measurements. Note
that using a median filter would not diminish the round-off effect. 

We also consider V-band light curves of comparable length obtained for
two Hubble-Sandage variables in M33, Romano's star (the light curve itself
is described in \citet{V532_photo}) and Hubble-Sandage Variable C
\citep{varc88}.

% table 1:
\begin{table}
\centering
\caption{ Objects selected for analysis. For Galactic objects, the
  numbers of data points are given after binning. }\label{tab:objs} % \label{tab:objs} }
\center{\small
\begin{tabular}{lccc}
Object ID & Variability Limits, mag & 
Time span & Number of Data Points \\
% & & giant eruptions & S~Dor loops & microvariability and
%long-timescale variations & & \\
\hline
\noalign{\smallskip}
 \multicolumn{4}{c}{Galactic objects} \\
\noalign{\smallskip}
% \astrobj{S~Dor}    & 8.0$\div$ 10.1 &  1976-12-23..2010-05-19 & 311\\
P~Cyg    & 4.5$\div$ 5.6  &  1917-08-13..2010-05-07 & 3396 \\
% \astrobj{HR~Car}   & 7.2$\div$ 9.3  &   1987-06-30..2010-06-04 & 396 \\
$\eta$~Car   & 4.5$\div$ 8.0  &  1943-07-24..2010-07-19 & 4637 \\
AG~Car   & 5.6$\div$ 8.9  &   1939-12-05..2010-05-03 & 2186 \\
%\astrobj{V840~Cyg} & 8.9$\div$ 10.3 &   1999-07-08..2010-06-13 & 30 \\
%\astrobj{V766~Cen} & 6.0$\div$ 8.0  &   1986-04-27..2010-06-25 & 906 \\
\noalign{\smallskip}
 \multicolumn{4}{c}{M33} \\
\noalign{\smallskip}
Romano's~star & 16.1$\div$ 18.6  &  1949-12-20..2009-11-08 & 802 \\
Var~C & 15.3$\div$ 17.9  &  1961-09-13..2005-11-08 & 635 \\
\noalign{\smallskip}
\end{tabular}
}
\end{table}

\subsection{Notes on Individual Objects}

\subsubsection*{P~Cyg}

P~Cyg is known for its outbursts in 1600 and later
\citep{degroot}. However, in the considered nearly 100-year time span it
hardly shows any signatures of activity, except for rare short-lasting
low-amplitude excursions \citep{markova01}. 
% In our sample it is unique in having no noticeable flaring
% activity. 
Variability amplitude is primarily contributed by
low-frequency variability. 
Several periodicities, from 17 to 100 days, were reported
\citep{ikolka}.
%, but none of
%them proved to be coherent. 

\subsubsection*{$\eta$~Car}

During the considered time span, its behaviour is as well relatively
quiet, especially if compared with its tremendous outbursts in the XIX
century. In our sample, $\eta$~Car is the only proven binary. Its
orbital period does not show up strongly in the optical
variability, probably because of the large optical depth of the outflows in the
optical. The variability pattern is dominated by the gradual increase
of the optical luminosity.
%, therefore the observed steep power spectrum
%may be interpreted as simply the spectrum this secular trend. 

\subsubsection*{AG~Car}

Variability of this prototypical LBV star was considered in
\citet{genderen_discoveries} together with that of $\eta$~Car. The
object exhibits several strong flares. Periodicity of $\sim 370^{\rm
  d}$ (somewhat smaller than the
characteristic flare duration time scale) is known to be present.

%\subsubsection*{\astrobj{V766~Cen}}

%Variations of this object during the period under consideration is
%similar to that of AG~Car. However, the object it not exactly of the
%LBV variability type but may be rather characterised as a ``luminous
%yellow variable'' ($\rho$~Cas type hypergiant of the spectral class
%F$\div$G). From all the six objects, it is the poorest example of an
%LBV. [consider a different fourth object??]

\subsubsection*{Romano's Star, or V532 (M33)}

One strong flare with an amplitude of about 2$\magdot{\,}$ and a
couple of smaller maxima represent a typical picture of S~Dor
variability. 
The V-band data and data reduction process are described
in \citet{V532_photo}. The object is hotter than all the other stars
of the sample
and reaches the spectral class of WN8 in its low/hot state
\citep{polcaro,us10}. Its spectrum in the quiet state (BIa$^+$e) is similar to
that of P~Cyg. 

\subsubsection*{Var C (M33)}

Behaviour of this variable denoted as Variable C by \citet{HS53} was
studied by \citet{varc88}, who conclude that the object behaves as a
typical S~Dor variable with a higher than average mass loss rate. The
variable spends about a half of its time in a high
and cool phase of the S~Dor cycle. Amplitudes of the flares reach
2$\magdot{.}$5 in the visual band. The data were provided by Alla
Zharova (private communication).

%%%%%%%%%%%%%%%%%%%%%%%%%%%%%%%%%%%%%%%%%%%%%%%%%%%%%%%%%%%%%%%%%%%%%

\section{Analysis and Results}\label{sec:AR}

All the time series are non-uniform but free from strong aliases,
therefore we use extirpolation method \citep{PR89}, where the initially
non-uniform time series is interpolated upon a regular grid, 
to compute the Fourier spectra. Fourier transform for uniform series was
performed using the {\tt fftw } library \citep{FFTW05}. Software
written in C and IDL was used for extirpolation and binning of the
power density spectra (PDS). PDS were binned by 10. We use relative
normalization defined as follows:

\begin{equation}
P_j = \frac{1}{\Delta T} \langle F \rangle^{-2} \left| \tilde{F}_j\right|^2
\end{equation}

where $\langle F\rangle$ is the mean flux, $ \tilde{F}_j$ is the
$j$-th component of digital Fourier transform of the flux $F$ (already
interpolated over a regular grid), $\Delta T= N \Delta t$ is the
total effective time span, $N$ is
the number of data points, $j=0..N-1$, $\Delta t$ is the
spacing of the regular grid. This normalisation has an evident
physical meaning of the relative variability power in a
unit frequency range. % For a uniform time series, i
Its expectation does not depend on the time span considered
and on data binning. Variability amplitude may be estimated by integrating the PDS
over frequency domain and taking a square root. 

\begin{figure*}
 \centering
\includegraphics[width=\textwidth]{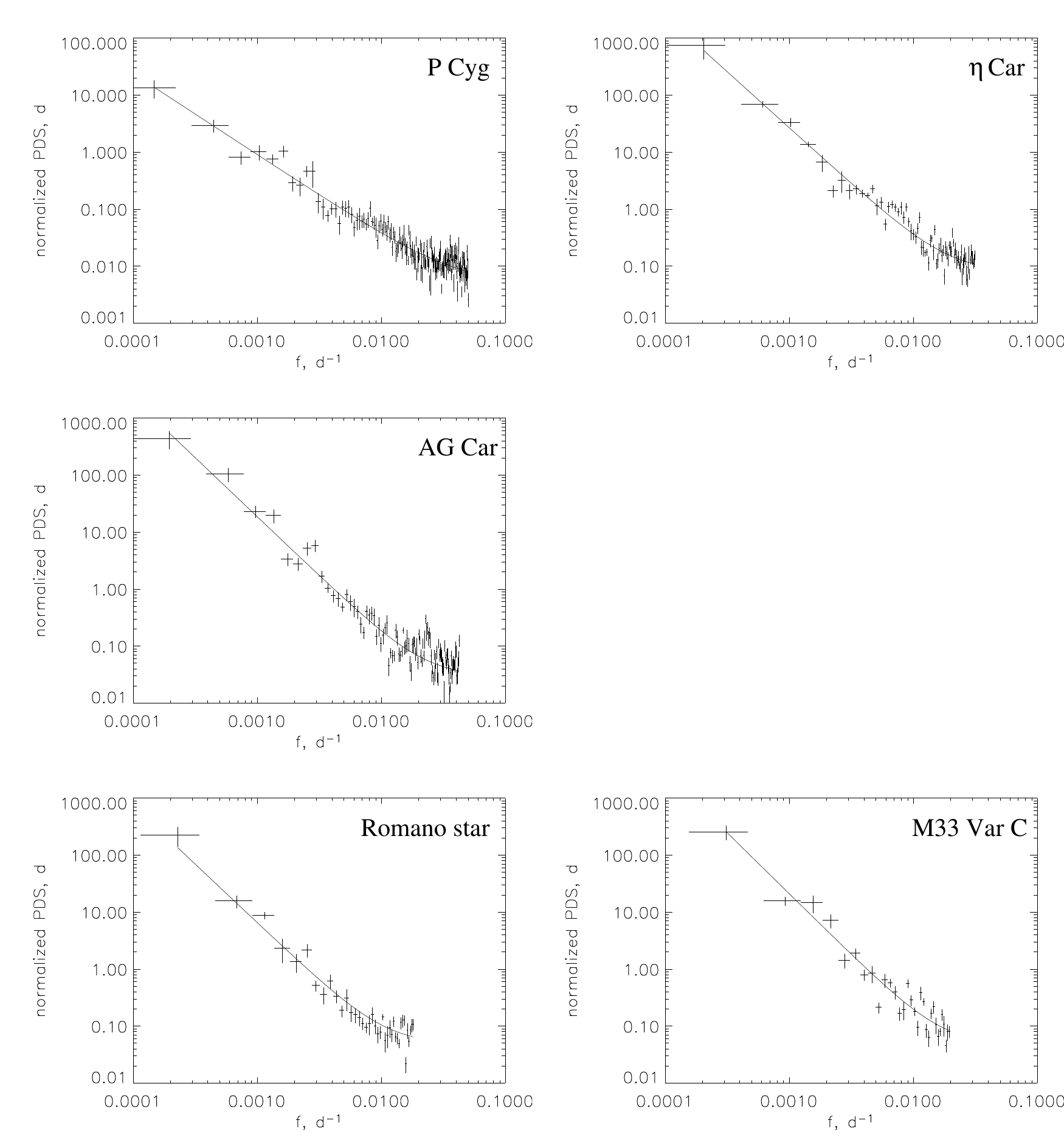}
\caption{Power density spectra of the five variables fitted by the power-law +
white noise model. }
\label{fig:pds}
\end{figure*}

All the obtained PDS (see figure \ref{fig:pds}) have approximately
power-law shapes with putative flattenings at lower ($\lesssim 10^{-4}\rm
d^{-1}$) and higher ($\gtrsim 0.01\rm d^{-1}$) frequencies. The
lower-frequency turnovers are smoothed out by the binning in the
frequency domain, but are better seen at finer binning. The data
were fitted with power-law ($PDS=N\cdot f^{-\alpha}$),
Lorentzian ($PDS=N/\left(1+\left(f/\gamma\right)^2\right)$, with zero
resonance frequency) and power-law
+ white noise ($PDS=N\cdot f^{-\alpha}+N_0$) models. Our choice of the
third model is justified by existence of the observational errors that create an
uncorrelated (white) noise component.
Lorentzian function is relevant because it is expected to represent the PDS
of a Poissonian sequence of exponentially decaying flares (and,
approximately, rapidly-decaying flares of other shapes; see section
\ref{sec:flaring}). It indeed proves to be a reasonable fit for the
objects that demonstrate strong flaring activity.
Fitting results are given is table \ref{tab:objfits}. 
%Usage of the last
%one is justified by the observational errors that create an
%uncorrelated (white) noise that co-adds with the observed ``red'' spectrum of
%the object variability. 

%table 2:

\begin{table} %\label{tab:objfits} 
\centering
\footnotesize
\caption{ PDS Fitting Results. }\label{tab:objfits} 
\begin{tabular}{p{2.5cm}ccccc}
 &  \multicolumn{3}{c}{Galactic} & \multicolumn{2}{c}{M33} \\
& P~Cyg  &  $\eta$~Car &  AG~Car  &  V532 &  Var~C\\
\hline
\noalign{\smallskip}\\
\noalign{Power Law: }\\
 $N$, $10^{-4}$d$^{-1}$  & $1.9\pm 0.2 $  &  $5.3\pm 0.4 $ & $1.6\pm
0.3 $  &  $0.59\pm 0.17 $  &   $0.42\pm 0.09 $\\

  $\alpha$ & $1.17\pm 0.02$ & $1.31\pm 0.01$ &  $1.56\pm 0.04$ & $1.63\pm
0.05$ & $1.86\pm 0.05$  \\ 

 $\chi^2/DOF$ &  $549/168$  &  $1393/230$  &  $545/107$   &  $166/38$
&  $213/30$ \\

\noalign{\bigskip}\\
\hline
\noalign{\smallskip}\\
\noalign{Lorentzian:}\\

$N$, d$^{-1}$  & $38\pm 4 $ & $770\pm 30 $ & $1900\pm 1200 $ &
$1.23\pm 1.07$ &   $1.51\pm 0.18 $ \\

$\gamma$,  $10^{-4}$d$^{-1}$ &  $2.16\pm 0.07 $ & $1.69\pm 0.03 $ &   $1.4\pm 1.3 $ &  $1.23\pm 1.07$ &  $1.51\pm 0.18 $ \\

$\chi^2/DOF$ &  $1602/168$ & $2574/230$ & $2167/107$ & $169/38$ & $227/30$ \\

\noalign{\bigskip}\\
\hline
\noalign{\smallskip}\\
\noalign{Power Law with White Noise Component:}\\

     $N$, $10^{-4}$d$^{-1}$  &   $0.47\pm 0.07 $ &  $0.67\pm 0.06$ &  $0.090\pm 0.007$ &   $0.008\pm 0.004 $ &  $0.08\pm 0.03 $ \\

$\alpha$ &  $1.43\pm 0.03$ & $1.75\pm 0.02$ & $2.12\pm
0.01$ & $2.34\pm 0.01$ & $2.14\pm 0.07$ \\

$N_W$, $10^{-4}$d$^{-1}$  & $45.2\pm 4.9 $ &  $139\pm 6 $ & $302\pm 15 $ & $550\pm 20$ & $500\pm 100 $ \\

 $\chi^2/DOF$  & $536/167$ & $1344/229$  & $455/106$ & $123/37$   & $199/29$ \\
%\hline
\end{tabular}
\normalsize
\end{table}

% figure 1

P~Cyg here is unique in its spectral shape that has the hardest
power-law slope spanning a frequency range of two orders of
magnitude. It is also unique in being the least active among all the
objects. Other stars that exhibit S~Dor cycles and outbursts have
steeper spectral slopes close to 2 with possible indications for a
low-frequency flattening. Average LBV power spectrum clearly has a
power-law shape between tens of days to years or tens of years and
possibly further toward hundreds of years. Gaining
statistics and increasing the time spans to hundreds of years may allow
to clearly resolve the turnover at about $f\sim 10^{-4}\,\rm
d^{-1}$ that definitely should take place because the light curve
should remain reasonably uniform in time. 

All the spectra exhibit curvatures and diffuse peaks that make all the
fits rather poor ($\chi^2/DOF \sim 2\div 5$). The overall broad-band shape
is, however, in good agreement with a single power law (with a
high-frequency contribution from observational uncertainties) in all
the cases.

\section{Variability Pattern of a Pseudo-Photosphere}\label{sec:varshape}

%Apparently there is no reason for LBV stars to change their bolometric
%luminosity on the time scales of S~Dor cycles. 
%Therefore, it is often
%Bolometric luminosity of LBV stars is often 
%proposed constant, while the optical variability is attributed to
%the variable size of the pseudo-photosphere of the wind. 
The idea of pseudo-photospheric nature of LBV cycles may be traced down
to the note of \citet{lamers87} who proposed that the primary source of the
variability in these objects is the variability of the mass-loss
rate. Below, we will assume that this variability is fast and
uncorrelated (white noise in mass loss rate) and show that such
variations indeed lead to a Brownian-like noise in the PDS, but with a
turnover at a higher frequency. Lower-frequency parts of LBV PDS require
some additional mechanism of stochastic variability. 

\subsection{Basic Assumptions}

Excluding brightest outbursts, variability of luminous blue
variables is generally assumed to be purely spectral. According to
\citet{lamers87}, the basic mechanism responsible for variable
luminosities of LBV stars in the optical is through variations in the mass
loss rate resulting in variable pseudo-photosphere radius and
effective temperature. %  Bolometric luminosity remains more or less constant. 
%It is usually assumed that the observed effective
%temperatures and radii of LBV stars indeed correspond to a moving
%pseudo-photosphere rather than some hydrostatic (on average)
%radius. 
%Therefore, variability of the photometric observables is best
%understood through the variability of the mass loss and velocity. 

In the relatively hot, relatively rarefied atmospheres of massive
stars opacities are dominated by
Thomson scattering, therefore the (radial) optical depth through the
wind may be calculated as:

%[ Thomson scattering dominates at higher temps! free-free has a 
%$\sim 10\div 100$ higher impact; however, only if the temperature
%  falls off as $\propto R^{-1/2}$. Reference to OPAL]

\begin{equation}\label{E:tau}
\tau(R, t) = \frac{\kappa_T}{4\pi} \int_R^{+\infty}
\frac{\dot{M}(t-r/v(r)) dr}{r^2 v(r)}
\end{equation}

Here, $v=v(r)$ is the velocity of
the wind. Different true absorption processes also contribute to the opacity of
the wind, but their effect depends on the temperature structure of the
wind and on the effects of clumping. We neglect all the effects
of true absorption though they are able to increase the opacity of the wind
by a factor of several. This will mimic a larger
mass accretion rate in the model.
 In its simplest form, the radius and effective temperature % (and
%therefore, a reasonable approximation for the broad-band spectrum)
 as a function of time may be estimated by equating the optical depth to
some fixed value (below we use $\tau=2/3$). 
Equation (\ref{E:tau}) may be then solved for $r$.

\subsection{Numerical Simulations}

Six model light curves (see table \ref{tab:sim:lc}) were calculated using a pseudo-photosphere
model with a blackbody photosphere defined by the condition
of $\tau=2/3$. Bolometric luminosity was everywhere set to $L_{bol}=3\times
10^{39}\ergl$ that is close to its value for P~Cyg
as estimated by \citet{PP90}. All the curves are 50 years in length
and contain between 195 and 9774 data points. Model F was aimed on reproducing the basic
variability properties of P~Cyg. Others differ in the unperturbed mass
loss rate $M_0$, mass-loss rate dispersion and wind velocity.

We use the semi-empirical $\beta$-law
$v(r)=\left(v_{\infty}-v_0\right)\left( 1-r/R\right)^\beta+v_0$ with
$\beta=1$. Wind acceleration spatial scale $R$ is assumed identical to
the hydrostatic inner radius
$R$, and we set $v_0=2.5 \kms$ everywhere.
Wind velocity at infinity was fixed, while the mass
loss rate experiences log-normal variations that we assume uncorrelated
on the time scales of interest. Of course, real stellar winds have
variable velocities, but simultaneous variations of wind velocity and
mass loss rate do not change the picture qualitatively. Variable wind
velocity would however lead to formally divergent solutions for
density distribution. In real winds, shock waves are expected to be
formed. Hence the effects of variable wind velocity should be
considered together with deviations from spherical symmetry and
quasi-stationarity. 
 Log-normality of the mass-loss rate distribution
is a useful assumption because is allows to
easily account for dramatic changes in the value of
$\dot{M}$ without producing unphysical negative values.
If the mass-loss rate varies on
 dynamical time scales, the wind will work as an integrator, making the
observed power density spectrum softer than the spectrum of $\dot{M}$
variations. For most of the models, we
assume the mass loss rate logarithm dispersion $\sigma\left(\ln
\dot{m}\right) = 2$ that corresponds to about a factor of $7$ change
in the mass loss rate itself and may account for the observed $\sim
1\div 2\magdot{\,}$ variations of S~Dor cycles. Models D and E
illustrate the effect of variable amplitude of mass loss variations. 

Inner (hydrostatic) radius was everywhere set to $2\times 10^{12}\rm
cm$, the scale height of the $\beta$ law for velocity is ascribed the
same value. 

% figure 2
\begin{figure*}
 \centering
\includegraphics[width=\textwidth]{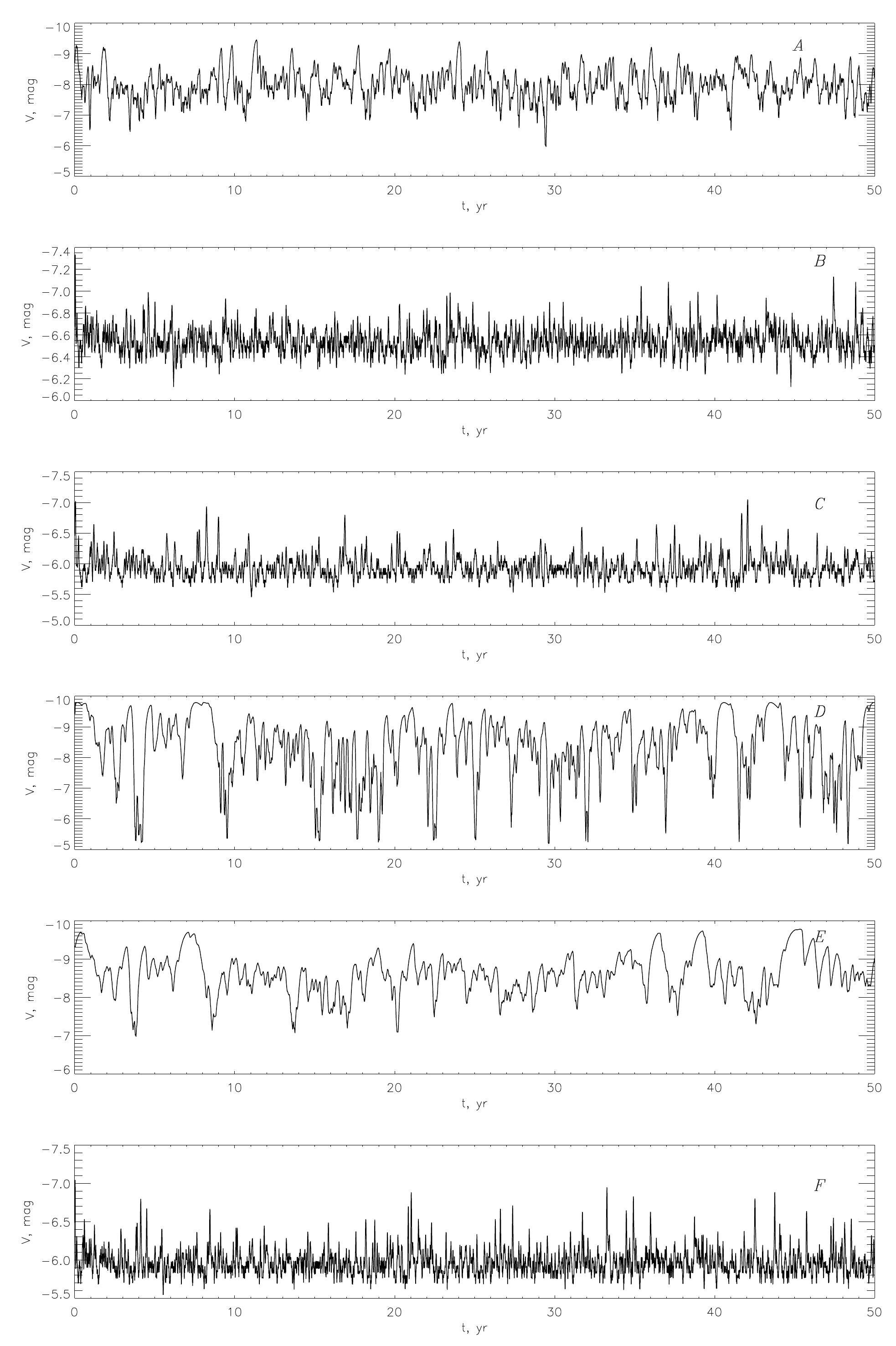}
\caption{
Simulated light curves. 
}
\label{fig:sim:lc}
\end{figure*}

\begin{figure*}
 \centering
\includegraphics[width=\textwidth]{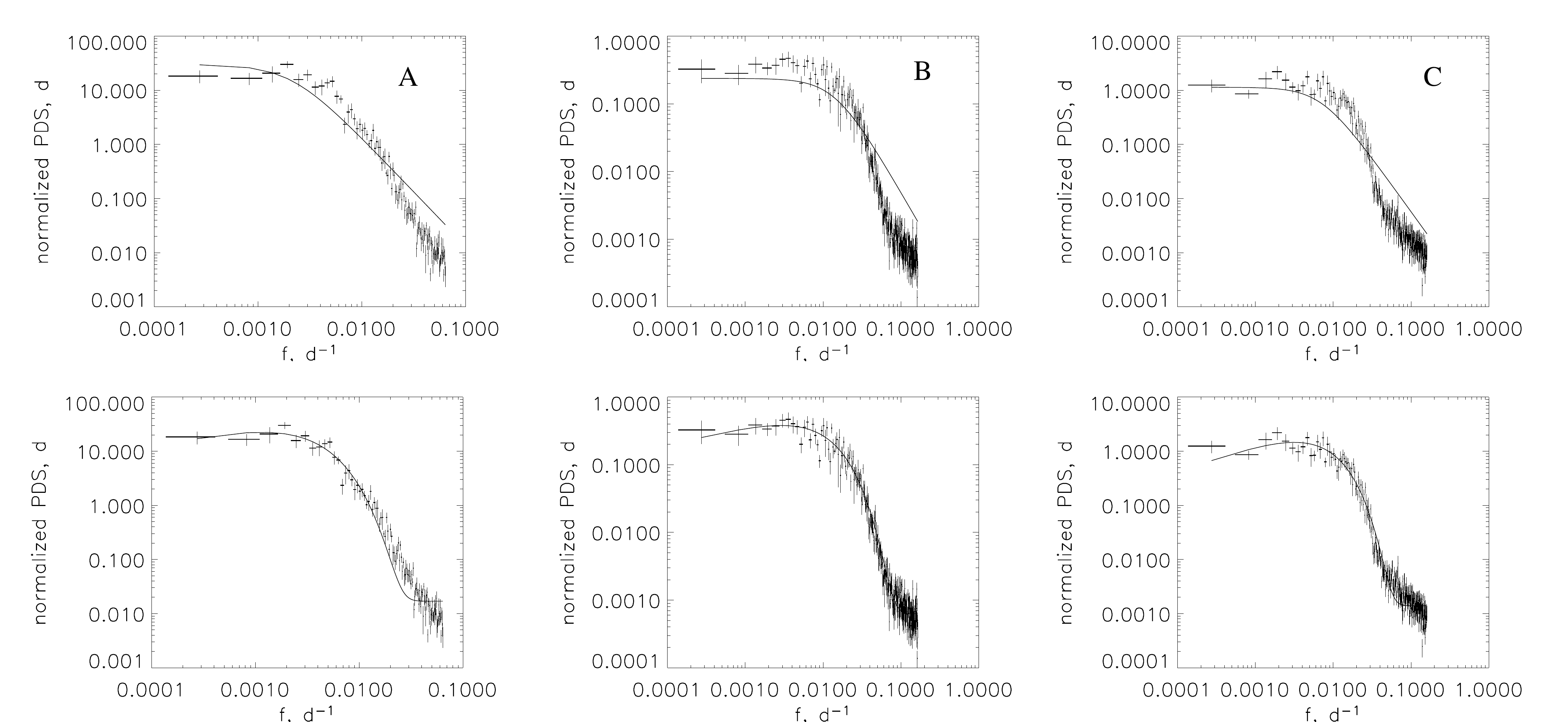}
\caption{
PDS fitting by Lorentzian (upper panels) and modified power law (lower
panels) models for simulated light curves, models A-C.
}
\label{fig:sim:pds1}
\end{figure*}

\begin{figure*}
 \centering
\includegraphics[width=\textwidth]{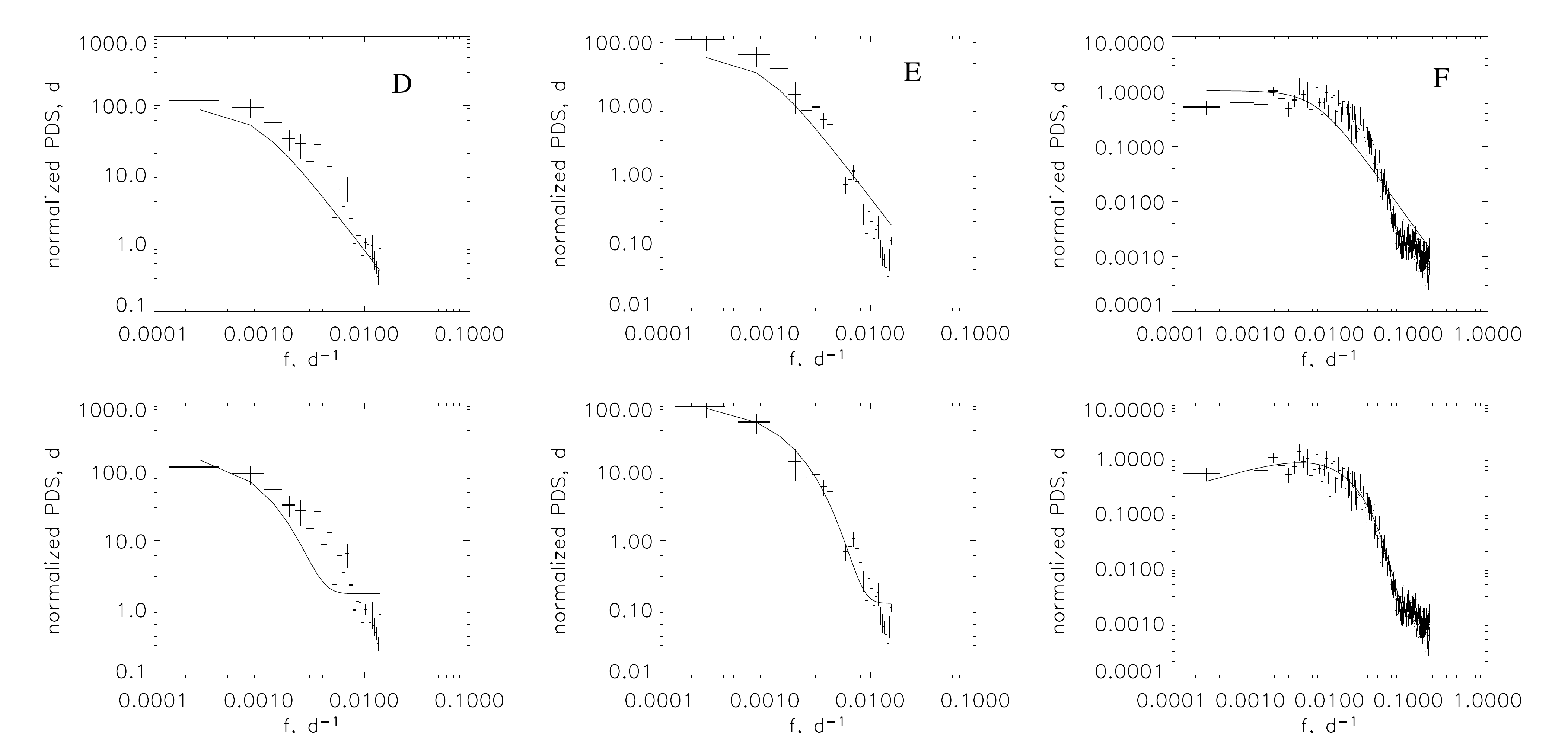}
\caption{
PDS fitting by Lorentzian (upper panels) and modified power-law (lower
panels) models for simulated light curves, models D-F.
}
\label{fig:sim:pds2}
\end{figure*}

In figure \ref{fig:sim:lc}, we show the simulated light curves in the
Johnson V band. Corresponding power density spectra are shown in
figures \ref{fig:sim:pds1} and \ref{fig:sim:pds2} together with
modified power law and Lorentzian
approximations. Modified power law is defined as follows:

\begin{equation}\label{E:modPL}
F_{MPL} = N f^{p}\exp(-T f)+N_0
\end{equation}

This spectral law differs from the {\tt power law + white noise} model used
above by an exponential decay factor with characteristic time $T$,
close to the wind replenishment time. 
 If the mass loss is uncorrelated at
higher frequencies, small fast variations should have a flat spectrum
as well (see section \ref{sec:linear}).

Basic parameters of the light curves and fitting results are
given in tables \ref{tab:sim:lc} and \ref{tab:sim:pds}, respectively. 
 $N(t_{dyn} < t_{dif})$ in the table
is the fraction of time when dynamical timescale defined as $t_{dyn} =
R_0/v$ is smaller than the radial diffusion time scale defined as $t_{diff} = \tau
R_0/c$, where $\tau$ is the total optical depth throughout the wind. If this
inequation is satisfied, the assumption of instantaneous energy
transfer vital for the pseudo-photosphere approximation
is violated (see section \ref{sec:disc:lvar}).
Power density spectra of the simulated light curves are shown in
figures~\ref{fig:sim:pds1} and \ref{fig:sim:pds2}

% table 3
\begin{table}
{\centering\footnotesize
\caption{ Light Curve Simulations }\label{tab:sim:lc} 
\begin{tabular}{lcccccc} %||p{3.5cm}|p{3.5cm}|p{3.5cm}}
\noalign{\smallskip}\\
Model ID &  A & B & C & D & E & F \\
\noalign{\smallskip}\\
\hline
\noalign{\smallskip}\\
 $\dot{M}_0$, $\Msunyr$ & $10^{-4}$ &  $10^{-4}$ & $10^{-5}$ & $10^{-4}$ & $10^{-4}$ & 
 $ 10^{-5}$ \\ 
 $\sigma\left(\ln \dot{m}\right)$ & 2 & 2 & 2 & 3.5 & 1 & 1 \\
$v/100\kms$ & 1 & 0.5 & 1 & 1 & 2 & 2 \\
 $R_0$, $10^{12}\rm cm$ & 28 & 14 & 3.7 & 28 & 55 & 3.7 \\
 No of points & 3910 &  1955  &  9774  &  195  &  1955  &  5585 \\
$M_V$ range, mag & -6.6..-10.0 & -6.7..-10.0 & -5.7..-8.2 & -6.1..-10.0 &
 -6.6..-10.0 & -5.7..-7.9 \\
$\langle M_V\rangle$, mag &   -8.2 & -9.0 &  -6.0 & -9.0 & -9.0 & -6.0
 \\
$\sigma\left( M_V\right)$, mag &  0.5 &  0.5 & 0.24 & 0.67 &  0.5 &
 0.22 \\
$N(t_{dyn} < t_{diff})$, \% &  0.2  &  0 & 0 & 4 & 0.2 & 0\\
\end{tabular}}
\end{table}

%table 4:

\begin{table}
\centering \footnotesize
\caption{ PDS of the Simulated Light Curves }\label{tab:sim:pds} 
\begin{tabular}{p{2.5cm}cccccc} 
\noalign{\smallskip}\\
%\hline
Model ID &  A & B & C & D & E & F \\
\noalign{\smallskip}\\
\hline
\noalign{\smallskip}\\
\noalign{Lorentzian:}\\
$N$, d  & $30\pm 1 $ & $0.468\pm 0.007 $ & $0.91\pm 0.09$ &  $120\pm
10 $ & $42\pm 2$ &  $1.2\pm 0.1 $ \\
$\gamma$, $10^{-3}$d$^{-1}$ & $1.44\pm 0.05 $ &  $3.92\pm 0.19$ &
$14\pm 3 $ &  $1.2\pm 0.3$ & $0.9\pm 0.1 $ &  $6.48\pm 0.05$ \\
$\chi^2/DOF$ &  $2527/115$ & $1304/298$ & $1299/291$ &  $55/24$ & $947/27$ & $1700/333$\\
\noalign{\bigskip}\\
\hline
\noalign{\smallskip}\\
\noalign{Modified Power Law:}\\
 $N$, d  & $550\pm 40$ &  $2\pm 4 $ & $40\pm 40$ &  $284\pm 3 $ & $104\pm 1$ & $12\pm 7 $ \\
$T$,  d &  $361\pm 8 $ &  $100\pm 28 $ &  $160\pm 20 $ & $614\pm 17 $ & $8.55\pm 5$ & $103\pm 11 $ \\
$p$ &  $0.41\pm 0.03$ &  $0.3\pm 0.2$ &  $0.46\pm 0.16$ &  $0.00\pm
0.01$ &  $0.00\pm 0.01$ &  $0.41\pm 0.10$ \\
$N_0$, d & $17\pm 9 $ &  $0.631\pm 2 $ & $1\pm 3 $ &  $0.6\pm 0.2 $ &
$110\pm 10 $ & $1\pm 2 $ \\
$\chi^2/DOF$  & $1161/113$ & $1128/296$  & $1443/289$ & $57/22$ & $173/25$ & $1362/331$\\
\end{tabular}
\normalsize
\end{table}

% figure 3

% figure 4

\subsection{Approximate Solution for Small Variations}\label{sec:linear}

For the linear pseudo-photosphere described by the equation
(\ref{E:tau}), it is easy to find the PDS for $\tau$ in the assumption
$v=const$. For small variations, $\tau$, $r(\tau)$ and the resulting
luminosity will have identical spectral shapes. 

Let $R_0$ be the radius corresponding to the photospheric radius
calculated for $\dot{M}_0$:

$$
R_0=\frac{3\kappa_T \dot{M}_0}{8\pi v}
$$

If $\dot{M}(t)$ is uncorrelated noise with the mean of $\dot{M}_0$ and
dispersion $\dot{M}_0 D\dot{m} \ll \dot{M}_0^2$, the resulting PDS will 
be identical to the spectrum of the equivalent $\delta$-function response:

$$
\tau_\delta= \frac{2}{3} \sqrt{D\dot{m}} \frac{1}{t^2}
$$

Here, $t>R_0/v$, because only visible matter influences the opacity. 
Fourier image:

$$
\begin{array}{l}
\tilde{\tau}(f) =  \frac{4\pi}{3} \sqrt{D\dot{m}} f \int_{2\pi f R_0 /
  v} \frac{e^{-i\eta}}{\eta^2} d\eta = \\ 
\qquad{} = \frac{4\pi}{3} \sqrt{D\dot{m}}
f \left( -\frac{\cos a}{a} - \si(a)+i \left(\frac{\sin a}{a} +
\ci(a)\right)\right), \\
\end{array}
$$

where integral sine and cosine functions (si and ci) are defined as integrals from
$a=2\pi f R_0 / v$ to positive infinity. Finally, power spectrum in
the linear case behaves as:

$$
\begin{array}{l}
PDS \propto D\dot{m} f^2 \left(a^{-2} + 2 a^{-1} (\cos a \si a +
\sin a  \ci a) + \ci^2 a +  \si^2 a \right) \propto\\
\qquad{ } \qquad{} \propto
\left\{
\begin{array}{lc}
f & f\ll v/2\pi R_0 \\
1 & f\gg v/2\pi R_0 \\
\end{array}
\right. \\
\end{array}
$$

%While first, $PDS\propto f$, asymptotics is confirmed by our
%simulations, at higher frequencies the approximation is violated
%because of the finite correlation time of mass loss variations.
The resulting PDS should be peaked, similar to the (\ref{E:modPL}) law
with $p=1$. The predicted power spectrum is vastly different from the
quiet-state variability spectrum of LBVs.

\subsection{PDS in the Flaring Case}\label{sec:flaring}

The strongly-variable \textbf{D} model has a
PDS qualitatively similar to those of strongly variable objects like
AG~Car. In fact, any source exhibiting smooth light variations
during individual uncorrelated flare events is expected to have flat
spectrum at the smallest frequencies and a Brownian-shaped $\propto
f^{-2}$ noise in the opposite limit ($f\to \infty$). If one considers
a flare of duration $t_f$ (smooth in its maximum but with rapid rise
and decay), its PDS shape in the
high-frequency limit may be estimated as:

$$
|\tilde{F}(f)|^2 \propto \left|\int_{-t_f/2}^{t_f/2}l(t) e^{-i 2\pi f
  t} dt\right|^2 \propto \frac{\sin^2\left( \pi f t_f\right)}{f^2}
\sim f^{-2}
$$

This approximation works for $f\gg 1/t_f$. In the opposite frequency limit,
variability is a Poissonian series of short events that implies white
noise. The spectral slope changes somewhere around $f\sim 1/t_f$, and
the exact shape of the knee in the PDS depends on the shape of a
single flare. The best known example is the Lorentzian spectrum for
exponential flares.

Qualitatively, this spectral shape is recognisable in objects like
AG~Car and Romano's star. The steep $\alpha \sim 2$ slope is probably
connected to the smooth shape of their flares. Yet the slope remains
steep in the lower-frequency domain, and
 the predicted turnover in the power spectrum has a frequency a couple of
orders of magnitude higher. This may be attributed to the existing
longer-time variability trends and to correlated behavior of outbursts and S~Dor
cycles. 

Though there are indications for periodicities such as the
one-year period in AG~Car, a slope holding for two orders
of magnitude requires some power-law scaling present. The
duration-amplitude relation reported by \citet{genderen_discoveries} may be the
key. Let us assume that the shape of a flare is determined by a single
parameter $t_f$ (its duration), but its amplitude is $\propto
t_f^p$. There are indications for $p\simeq 1$. A Poissonian series of
such events will be characterised by the following power spectrum:

$$
\left|\tilde{F}(f)\right|^2 \propto f^{-2} \left| \sum t_{f, i}^p
\sin(\pi f t_{f,i})  \right|^2\simeq f^{-2(1+p)} \left|\int x^p
sin(\pi x) n(x/f) dx \right|^2 
$$

Here, $n(t_f) = dN/dt_f$ is distribution of flares per unit duration
time. If $p\simeq 1$, distribution of the form $n(t_f) \propto t_f^{-1}$ (uniform if
one considers $\ln t_f$ interval) may explain
the observed spectral slopes. 

\section{Discussion}\label{sec:disc}

\subsection{Different Kinds of Variability?}

One principal question about LBV variability is whether
micro-variability (with temporal scales less than several months and amplitudes
$\lesssim 0\magdot{.}2$), S~Dor cycles, slow luminosity variations 
and giant eruptions are produced by a
single physical mechanism. As we show in this
article, the basic properties of LBV variability at shorter timescales
(up to $\sim 1\yr$) may be reproduced by
the toy model described in the previous section. Complex shapes of PDS
at frequencies $f\sim 10^{-3}\div 10^{-2}\rm d^{-1}$ (approximately
corresponding to the duration
timescale of an average flare) are possibly connected to transition
from the wind-smoothed pseudo-photospheric noise at higher frequencies to a
very similar Brownian noise at lower frequencies, probably connected
to slow pulsations or some other internal
processes. 

Pulsations studied by the hydrodynamical modeling in
\citet{fadeyev} are similar to the observed micro-variations in
periods and amplitudes. They are also a good candidate for the process
driving mass loss variations. Viewed at $\sim 1\yr$ timescales, irregular
pulsations are both fast and uncorrelated, that justifies the usage of
white noise approximation. Longer-timescale variations may appear if
some instabilities lead to secular evolution of LBV pulsations
\citep{buchler93}. 

Pseudo-photosphere approach is able to reproduce some of the key
points of LBV variability such as the steep power spectra. Even if
the higher opacity of the matter is taken into account, it is short to
explain the low-frequency part of the PDSs. This is consistent with the
conclusion of \citet{koter} who find that the observed variability of
LBV stars is too strong to be explained solely by velocity and
density variations in the wind. 

\subsection{Variable Bolometric Luminosity}\label{sec:disc:lvar}

Strongest deviations from this simple picture are the giant eruptions and ``supernova
impostors'' (see \citet{snimpostors} for review) characterised by significant changes in bolometric
luminosity (by two magnitudes and more, see for example
\citet{clark09}). Variable bolometric luminosities are impossible to
explain in the assumptions of the pseudo-photosphere approach. 

Note however that variable bolometric luminosity may still arise from
variations in mass loss rate if the characteristic diffusion
timescale in the wind becomes large enough. Generally, the
replenishment timescale (\ref{E:tref}) is much longer than
the radiation diffusion timescale that may be roughly estimated as $t_{diff}
\sim \tau R/c$. This makes the expanding atmosphere practically
quasi-stationary from the point of view of energy transfer. 
However, if the total optical depth in the wind becomes significantly larger than
the inverse wind velocity in $c$ units, $\tau > c/v \sim 10^3$, the energy
output from the pseudo-photosphere will be strongly modulated by the
envelope evolution. Adiabatic losses should be taken into account as
well in this case. 

%It should be noted that i
%If the above explanation is true there should
In these energy output modulation, there should 
be both increases and decreases of bolometric luminosity during the
outburst. The mean level of energy release in the stellar interior is
expected to be constant at much larger (thermal and nuclear) timescales of
$\gtrsim 10^4\yr$.% Besides this, b
Bolometric luminosity should 
become \emph{ smaller} during giant eruptions because of adiabatic
losses close to the photon-tiring limit \citep{photon_tiring}. During
strong flares, deviations from the mean luminosity level should hold on
timescales smaller than the survival time of an optically-thick
outburst of the mass $M$ expelled at the maximal velocity of $v_{out}$:

\begin{equation}
t_{out} \sim \frac{1}{v_{out}} \sqrt{\frac{\varkappa_T M_{out}}{4\pi}}
\sim 20\, \left(\frac{v_{out}}{100\,\kms}\right)^{-1}
\left(\frac{M_{out}}{1\,\Msun}\right)^{1/2} \yr
\end{equation}

This estimate naturally arises in a gas that remains hot. Changes in
opacity and in particular gas recombination produce a variety of
similar time scales proposed for the photosphere phase or
plateau stage durations in supernova light curves 
(see for example \citet{SNlc} and references therein).

Summarizing, we conclude that LBV eruptions should be accompanied
by additional energy input inside the hydrostatic radius. 

\section{Conclusions}

Variability of LBV stars is generally consistent with a single steep
power law, but the processes forming its low- and high-frequency parts
are probably different. The XX-th century light curve of P~Cyg does not
show any noticeable flares but is correlated at long timescales up to
decades. The overall power spectrum is consistent with a
$p=1.3\div1.4$ power law. 
For all the flaring sources, % including the YBV V766~Cen, 
the slope is higher, $p\simeq 2$. 

Emergence of this steep power-law spectral shape is reasonable in the
pseudo-photosphere approach. Short-timescale variations in mass loss
rate are effectively smeared and integrated over the line of sight
that leads to a Brownian noise with a break at the ``replenishment''
time of about one year. Smaller amplitude variability should produce
peaked noise with the maximum at several months. 
The resulting variability lacks two important
features of the real light curves: it is not correlated at 
longer timescales (slow variability
component) and conserves bolometric luminosity.  

Longer-timescale correlations may arise partially from the higher
opacity of the outflowing matter (that affects the observed luminosity
changes in a complex way not implemented in our model), variable
wind velocity or some internal mechanisms. 
% Slow pulsations and other internal processes are also
% likely to make contribution. 
%Bolometric luminosity variations may
%arise during intense eruptions when radiation becomes trapped in the
%stellar envelope. [ estimate the effect! ]

\section*{Acknowledgements}

Our article makes use of the AAVSO International Database, hence we
would like to thank all the AAVSO observers who made the present study
possible. Author also thanks Alla Zharova for providing the V-band
light curves of the two objects in M33.

\bibliographystyle{model2-names}
\bibliography{mybib}

\end{document}